\documentclass[aps,prb,twocolumn,showpacs,showkeys]{revtex4}

\usepackage[]{amsmath}
\usepackage{epsf}
\usepackage{psfrag}
\usepackage{mathrsfs}
\usepackage{graphicx}
\usepackage{dcolumn}
\usepackage{bm}

\begin{document}
\title{Non-equilibrium superconductivity in superconducting resonators}
\author{D. J. Goldie and S. Withington}
\affiliation {Detector and Optical Physics Group, Cavendish Laboratory, University of Cambridge,
J J Thomson Avenue, Cambridge, CB3 0HE, UK \\
}
\email{d.j.goldie@mrao.cam.ac.uk}

%
\date{\today}
%

\begin{abstract}
We have calculated the non-equilibrium quasiparticle and phonon distributions
$f(E)$, $n(\Omega)$, where $E$ and $\Omega$ are the
quasiparticle and phonon energies respectively,  generated by the
photons of the  probe signal of a low temperature superconducting resonator SR
operating well-below its transition temperature $T_c$ as the absorbed probe power per unit volume  $P_{abs}$ was changed. The calculations
 give insight into a rate equation estimate which suggests that the quasiparticle
 distributions
can be driven far from the thermal equilibrium value for typical
readout powers.
From $f(E)$
the driven  quasiparticle number density $N_{qp}$ and  lifetime $\tau_r$ were calculated.
Using $N_{qp}$
we defined an effective temperature $T_N^*$ to describe the driven $f(E)$.
The lifetime was compared to the distribution averaged thermal lifetime at $T_N^*$ and
good agreement was found typically within a few percent.
We used $f(E)$  to model a representative SR.
The complex conductivity and hence the frequency
dependence of the
experimentally measured
 forward scattering parameter $S_{21}$ of the SR as a function of  $P_{abs}$ were found.
The non-equilibrium  $S_{21}$ cannot be accurately modeled by a thermal distribution
at an elevated temperature $T_{21}^*$ having  a higher quality-factor in all
cases studied and
for low $P_{abs}$ $T_{21}^*\sim T_N^*$.
 Using
$\tau_r$ and $N_{qp}$ we  determined the
achievable Noise Equivalent Power of the resonator used as a detector as a function of $P_{abs}$.
Simpler   expressions for $T_N^*$ as a function of
$P_{abs}$ were derived which give a very good account of  $T_N^*$ and also $N_{qp}$ and $\tau_r$.
We conclude that multiple photon absorption from the probe increases
the quasiparticle number above the thermal background and ultimately limits the achievable NEP of the resonator.
\end{abstract}
%
\maketitle
\section{Introduction }
\label{sec:Intro}
Superconducting resonators (SRs) with high quality factor $Q$ operating at low reduced temperatures $T/T_c\simeq 0.1$, where $T_c$ is the superconducting
transition temperature, are used not only
as ultra-sensitive   detectors of individual quanta or incident power
for applications in sub-millimeter, millimeter, optical, X- and $\gamma$-ray astrophysics.\cite{Jonas_nature, Jonas_review, George_kid, Monfardini, Baselmans_review}
but also
as elements of Qubits for
quantum computing,\cite{Dicarlo_nature, Hofheinz_nature, Schoelkopf_nature}
They are also  needed as  elements of microwave Superconducting Quantum Interference Device (SQUID) multiplexers\cite{Irwin_SQUID_MUX}
and also  some SQUID geometries more directly.
Despite this technological importance, we are unaware of any detailed analysis of the effect of the interaction of a {\it flux} of microwave photons of frequency $\nu_p\ll 2\Delta(T)/h$, where $2\Delta(T)$ is the temperature-dependent  superconducting energy gap
and $h$ is Planck's constant,
on the superconducting state at low temperatures.
One might assume that if $h\nu_p< 2\Delta$ the photon interaction cannot change the quasiparticle number since
 the photon cannot break a Cooper pair.
Whilst true at the single quantum level, that assumption  ignores the effect of a  flux of photons comprising very
many quanta which might be used in a
typical experiment and the competing effects of multiple photon absorption and quasiparticle scattering and recombination on the driven quasiparticle
distribution $f(E)$ where $E$ is the energy.
Understanding the effect of non-equilibrium quasiparticles on a Qubit is certainly
a topic of current interest,\cite{Lenander_noneq_prb_2011, Martinis_prl_energy_decay}
where non-equilibrium quasiparticles  may be a limiting factor  on Qubit energy relaxation times.
Our interest is particularly in the context of non-equilibria in SRs used as quantum and power detectors
 although we would emphasize that our calculations
apply to SRs in general and the mechanisms and solutions we describe are common to all SR applications.
Indeed  the results presented below are  {\it independent} of the particular geometry or application
provided the effect of geometry on the power absorbtion is considered.

 In the context of power detection, SRs are sensitive to changes in detected power because of the dependence of the surface impedance $Z_s$ of the superconductor
on quasiparticle density and the dependence follows from the complex conductivity $\sigma$
described by Mattis and Bardeen.\cite{Mattis_Bardeen}
The SR  is embedded in an electrical readout circuit and is driven by  a
microwave probe signal of frequency $\nu_p \ll 2\Delta(T)/h$ close to the circuit resonant frequency $\nu_0$.
 If the quasiparticle density is changed, for example by absorption of a high frequency photon signal
of sufficient energy to break Cooper pairs $h\nu_{\Phi}>2\Delta$, where $\nu_{\Phi}$
is the detected photon frequency, $Z_s$ is changed  and the change can be monitored by measuring
the change in the  (complex) resonance transmission characteristic
$S_{21}(\nu)$  of the  probe signal. The change relaxes back to the unperturbed state as energy
is exchanged between the quasiparticles and the phonons of the superconductor and ultimately the substrate.
In this way very sensitive
power detectors can  be  made. For example Ref.~(\onlinecite{deVisser_apl_2012}) estimated
a dark   Noise Equivalent Power (NEP), i.e. ignoring the achieved signal detection efficiency,  of $2\times10^{-19}\,\,{\rm W/\sqrt{Hz}}$ at lowest readout power
 which was accounted for  in terms of the generation-recombination noise of the quasiparticles and a limiting lifetime of
$3-4\,\,{\rm ms}$.
  It has been predicted that NEP's approaching $10^{-20}\,\,{\rm W/\sqrt{Hz}}$ or lower may be achieved using SRs.
The readout naturally lends itself to frequency-division multiplexing where a large number of SRs each operating with a slightly different
$\nu_0$ are coupled to a through-transmission line.
A high $Q$ resonator  can be formed by  lithographically  patterning  a low-$T_c$
superconducting thin film such as Al ($T_c=1.2\,\,{\rm K}$) on a dielectric
substrate. The substrate is held at the bath temperature $T=T_b$.
A number of SR geometries are possible including ring, half- and quarter-wave or lumped-element designs.

In a recent paper we discussed the effect of the probe power on the resonator characteristic.\cite{de_Visser}
The key point
is that the readout is dissipative\cite{Thompson_2012, Jonas_review, deVisser_apl_2012} and the absorbed power
can be calculated knowing  the embedding electrical circuit.
Probe photons are absorbed by the quasiparticles of the SR
which  changes  $f(E)$, although  the resulting distribution has not yet been calculated.
In a real device other mechanisms may contribute to the dissipation for example dielectric or radiative losses.\cite{Jonas_review}
Here we focus on  the dissipation associated with the real part of $\sigma$.
The aim of this  work is  to calculate  the effect of a
microwave drive at low temperatures ($T_b/T_c=0.1$)
on the  static,
non-equilibrium quasiparticle and phonon
distributions in a superconductor as the probe-power levels are changed. 
 We also derive simpler analytical expressions which give
 a good approximation to the key results of the full calculation.

Energy relaxation processes of quasiparticles in a superconductor that couple to   phonons comprise
scattering of quasiparticles with absorption or emission of phonons, and
pair-breaking and recombination  of quasiparticles with absorption or
emission of phonons of energy $\Omega\ge 2\Delta$. Energy escapes from the superconductor as phonons enter the substrate.
 The coupled kinetic equations that describe these interacting subsystems were derived by Bardeen, Rickayzen and Tewordt\cite{Bardeen_Rickayzen}
 and
discussed in detail by Chang and Scalapino.\cite{Chang_and_Scalapino_prb,Chang_and_Scalapino_ltp}
In Ref.~\onlinecite{Chang_and_Scalapino_prb} the coupled equations were linearized and solved for a variety of
drive sources including microwaves. In Ref.~\onlinecite{Chang_and_Scalapino_ltp}
full non-linear
solutions were obtained. Crucially however those solutions were obtained close to $T_c$ where
microwave drive can lead to gap-enhancement effects. The  kinetic equations have been used  to investigate the non-linear
effect of high energy photon interactions at low temperatures.\cite{Zehnder_model, Ishibashi, Kurakado}
To our knowledge no full non-linear calculations exist of the effect of microwave drive at
low temperatures on SRs.

The structure of this paper is as follows. In Section~\ref{sec:Equilbrium estimate} we give an estimate of the
power densities  where non-equilibrium effects are likely to become important in a SR.
In Section~\ref{sec:Non-Equilbrium} we discuss the general properties of the coupled kinetic equations and derive the
form for the drive and subsystem power-flow  terms necessary to ensure energy conservation whilst also
discussing the effects of out-diffusion in a real geometry and the method used to calculate distribution averaged
recombination times.
In Section~\ref{sec:Numerical_parameters} we give numerical parameters used in calculations to describe a
clean thin-film superconductor (Al), and in
~\ref{sec:KID_model} we describe the model used to calculate $S_{21}$ for
 a representative SR.
  Section~\ref{sec:Method} describes  the numerical method.
In Section~\ref{sec:Non-Equilbrium_solutions} we show solutions for the non-equilibrium quasiparticle and phonon
distributions of a driven SR operating with $T_b/T_c=0.1$ as a function of $P_{abs}$
including calculations of driven quasiparticle density $N_{qp}$,
an effective temperature $T_N^*$ determined from  $N_{qp}$, and of the
distribution-averaged relaxation time $\tau_r$.
We
 calculate the driven $S_{21}$ for a
representative Al SR under the same powers. We also estimate the effect on the achievable NEP  for a
quantum SR detector using these results.
Section~\ref{sec:Model} describes two  models both giving a reasonable but simpler account  of the results.
In Section~\ref{sec:Discussion} we discuss the implications of the work with concluding remarks.
\section{Equilibrium estimate}
\label{sec:Equilbrium estimate}
In this section  we use the equilibrium interaction times derived by Kaplan {\it et al.}\cite{Kaplan} to estimate the power
densities where non-equilibrium effects are likely to occur in a SR at low temperatures.
Ref.~\onlinecite{Kaplan}
gives expressions for the thermal equilibrium lifetimes:
$\tau_s(E,T)$ for the
scattering of quasiparticles of energy $E$, $\tau_r(E,T)$ for recombination to Cooper pairs,
$\tau_{\phi s}(\Omega,T)$  for the scattering of phonons of energy $\Omega$ and
$\tau_{pb}(\Omega,T)$ for a phonon to break a pair. These are calculated in terms
of  characteristic quasiparticle and phonon lifetimes $\tau_0$ and $\tau_0^\phi$.
The volume density of thermal quasiparticles
is
$N(T) = 4 N(0)\int_\Delta^\infty  \rho(E,\Delta(T)) f(E, T) dE$,
where $\rho(E,\Delta)=E/\sqrt{E^2-\Delta^2}$ is the normalized quasiparticle density of states, $f(E,T)=\left(\exp\left(E/k_bT\right)+1\right)^{-1}$
 is the thermal Fermi distribution at temperature $T$ (in contrast to $f(E)$ the driven distribution),
  $k_b$ is Boltzmann's constant and $N(0)$ is the single-spin density of states at the Fermi energy.
Consider the case of an Al resonator  with  a probe signal
of  $\nu_p\sim 4\,\,{\rm GHz}$.
Since $h\nu_p\simeq 2\Delta/20$  it is not clear that
the probe-signal is capable of changing the equilibrium 
$N(T)$ and hence $\Delta(T)$.
Our measurements and modeling show that  typical experimental  read-out powers
 dissipate  of order $1 \to 100\,\, {\rm aW/\mu m^3}$ in  an Al SR
 used as a power detector.\cite{George_kid, de_Visser}

The probe photons are absorbed by the quasiparticles changing  their energy distribution $f(E)$.
 We expect the appearance of  at least one peak
 in $f(E)$  around $E=h\nu_p$
due to
 absorption of the monochromatic probe photons by the large density of quasiparticles near to the gap, perhaps more peaks if
  $P_{abs}$ is sufficiently high.
Assume that energy relaxation of an excited quasiparticle can only occur by scattering
with emission of phonons and
all of the emitted  phonons are lost from the film with an energy-independent time
$\tau_{l}$. Indeed, for thermal distributions at low reduced temperatures quasiparticle-scattering times are significantly shorter than
recombination times.\cite{Kaplan}
However if $f(E)$ has {\it any} non-equilibrium quasiparticles above $E=3\Delta$
 the phonon  emitted in scattering may have $\Omega>2\Delta$ which can
break a pair. This is the onset of non-equilibrium effects
since
the number of quasiparticles can be changed.
   \begin{figure}
   \begin{center}
   \begin{tabular}{c}
   \psfrag{yaxis} [] [] { $E_\delta/\Delta$  }
  \psfrag{Yaxis2} [] [] {  }
  \psfrag{xaxis} [] [] {  $P_{abs}\,\,(\rm {aW/\mu m^3}) $  }
   \psfrag{T1} [] [] {  $T/T_c=0.1$  }
     \psfrag{T2} [] [] {  $T/T_c=0.2$  }
     \psfrag{T3} [] [] {  $T/T_c=0.3$  }
   \includegraphics[height=8.6cm, angle=-90]{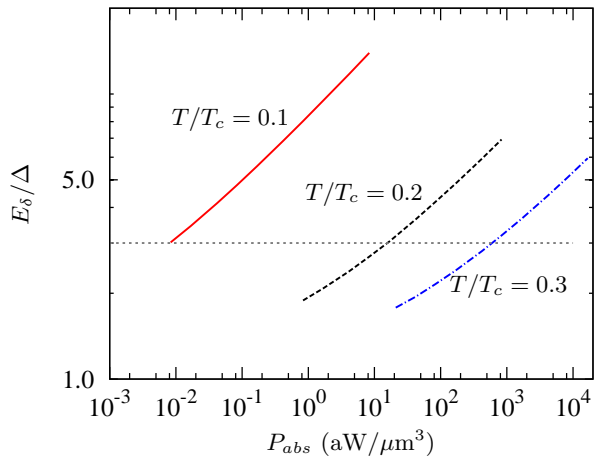}
   \end{tabular}
   \end{center}
   \caption[Fig1]
   { \label{fig:Ebar_taus}
(Color online) Quasiparticle energy $E_\delta$ as a function of $P_{abs}$ for an Al film at
  3 bath temperatures assuming
cooling by quasiparticle-phonon scattering alone.
 }
   \end{figure}
We can obtain a naive estimate of
the probe-power level for this to occur by assuming that all of the available quasiparticles at $T_b$
 i.e.  $N(T)\equiv N(T_b)$, are
driven to the same unknown energy $E_\delta$ and that scattering is the only energy loss mechanism for the quasiparticles,
so that $P_{abs}=E_\delta N(T_b)/\tau_s(E_\delta,T_b)$.
Figure~\ref{fig:Ebar_taus}
shows $E_\delta$ for three bath temperatures as a function of $P_{abs}$
calculated for an Al film with
 $\Delta(0)=180\,\,{\rm \mu eV}$, $T_c=1.17\,\,{\rm K}$,
$N(0)=1.74\times 10^{4} \,\,{\rm \mu eV^{-1}\, \mu m^{-3}}$,
and  $\tau_0=438\,\,{\rm ns}$.
At a typical operating temperature $T_b/T_c=0.1$ all the quasiparticles may have
$E_\delta >3\Delta$.
Even at  $T_b/T_c=0.2$  $E_\delta$ is close to the pair-breaking threshold  due to reabsorption of the scattered phonon for typical SR drive powers.
We emphasize that this estimate is very conservative because non-equilibrium effects occur as soon as there
are any non-thermal quasiparticles with $E \ge 3\Delta$ and the degree of non-equilibrium depends on the
probability of pair-breaking before loss by the scattered phonons, $\tau_{l}/\tau_{pb}$. At low temperatures
$\tau_{pb}=\tau_0^\phi$.
For a 100~nm
Al film on sapphire or Si we estimate that $\tau_{l}\sim \tau_{pb}.$\cite{Kaplan_loss}
This suggests that  the probe-signal may
break Cooper pairs in the driven SR even for very low $P_{abs}$ and scattering without pair-breaking
 may not be
 a sufficient energy-relaxation mechanism. Then a full
non-equilibrium description is required and we will not be surprised if the
power densities to observe non-equilibrium effects may be lower than suggested by Fig.~\ref{fig:Ebar_taus}.
\section{Non-Equilibrium Resonators}
\label{sec:Non-Equilbrium}
In this and subsequent Sections we describe and use a full non-linear
 solution of the kinetic equations for the coupled
 quasiparticle and phonon systems. The rates of change of the distribution functions, $f(E)$ for the quasiparticles and $n(\Omega)$
for the phonons,
 are given in
Eqs. (7) and (8) of Ref.~\onlinecite{Chang_and_Scalapino_ltp}. Substituting
 $\tau_0$ and $\tau_0^\phi$ these become
\begin{widetext}
\begin{equation}\begin{split}
\frac {df(E)}{dt} = I_{qp}(E,\nu_p)
 - &  \frac{1}{\tau_0(k_bT_c)^3}
\int_0^\infty d\Omega \Omega^2 \rho(E+\Omega)\left(1-\frac{\Delta^2}{E\left( E+\Omega \right)} \right) \times  \\
& \left( f\left(E\right) \left[ 1-f\left(E+\Omega\right)\right] n\left(\Omega\right) -
  \left[1-f\left(E\right)\right]f\left(E+\Omega\right)\left[n\left(\Omega\right)+1\right]  \right) \\
 - &   \frac{1}{\tau_0(k_bT_c)^3}
\int_0^{E-\Delta} d\Omega \Omega^2 \rho(E-\Omega)\left(1-\frac{\Delta^2}{E\left( E-\Omega \right)} \right) \times\\
& \left( f\left(E\right) \left[ 1-f\left(E-\Omega\right)\right] \left[ n\left(\Omega\right) +1\right] -
  \left[1-f\left(E\right)\right]f\left(E-\Omega\right) n\left(\Omega\right)  \right) \\
 - &   \frac{1}{\tau_0(k_bT_c)^3}
\int_{E+\Delta}^\infty d\Omega \Omega^2 \rho(\Omega-E)\left(1+\frac{\Delta^2}{E\left( \Omega - E \right)} \right) \times\\
& \left( f\left(E\right)  f\left(\Omega-E\right) \left[ n\left(\Omega\right)+1\right]  -
  \left[1-f\left(E\right)\right]\left[1-f\left(\Omega-E\right)\right] n\left(\Omega\right)  \right) ,
\label{Eq:C_and_S_1}
\end{split}
\end{equation}
and
\begin{equation}\begin{split}
\frac {dn(\Omega)}{dt} = - & \frac{2}{\pi\tau_0^\phi \Delta(0)}
\int_\Delta^\infty dE \rho(E)\rho(E+\Omega)\left(1-\frac{\Delta^2}{E\left( E+\Omega \right)} \right) \times \\
& \left( f\left(E\right)\left[1-f\left(E+\Omega\right)\right]n\left(\Omega\right)
-\left[1-f\left(E\right)\right]f\left(E+\Omega\right)\left[n\left(\Omega\right)+1\right] \right)\\
- & \frac{1}{\pi\tau_0^\phi \Delta(0)}
\int_\Delta^{\Omega-\Delta} dE \rho(E)\rho(\Omega-E)\left(1+\frac{\Delta^2}{E\left(\Omega-E \right)}\right) \times \\
& \left( \left[1- f\left(E\right)\right] \left[1-f\left(\Omega-E\right)\right]n\left(\Omega\right)
-f\left(E\right)f\left(\Omega-E\right)\left[n\left(\Omega\right)+1\right] \right) \\
- & \frac{ n(\Omega)-n(\Omega,T_b)}{\tau_{l}} ,
\label{Eq:C_and_S_2}
\end{split}
\end{equation}
\end{widetext}
where $n(\Omega,T_b)=\left(\exp(\Omega/k_b T_b)-1\right)^{-1}$ is the Bose distribution evaluated
at $T_b$. The
 term $I_{qp}(E,\nu_p)$ is the source term due to the photons at energy $E$ and quantifies the drive of the microwave probe.
The energy gap is modified from its equilibrium value and is determined self-consistently
so that
\begin{equation}
\frac{1}{N(0)V_{BCS}}=\int_\Delta^\infty dE \frac{1-2f(E)}{\sqrt{E^2-\Delta^2}} ,
\label{Eq:Non_eq_gap}
\end{equation}
where $V_{BCS}$ is the BCS interaction parameter.
 Eq.~(\ref{Eq:Non_eq_gap}) calculates the non-equilibrium $\Delta$
 using the non-equilibrium $f(E)$. In the static, driven situation to be solved
 $df(E)/dt=d n(\Omega)/dt=0$.
\subsection{Microwave drive term}
\label{subsec:Drive_term}
The form of  $I_{qp}(E,h\nu_p)$ was calculated by Eliashberg\cite{Eliashberg_1}
 and
\begin{widetext}
\begin{equation}\begin{split}
I_{qp}(E,\nu_p) =    2 B   \Bigg[ \rho(E+h\nu_p,\Delta)
& \left[ 1+ \frac {\Delta^2}{E\left(E+h\nu_p\right)} \right]
\left[f\left(E+h\nu_p\right)-f\left(E\right)\right] \\
  -  \rho(E-h\nu_p,\Delta)
& \left[1 +\frac {\Delta^2}{E\left(E-h\nu_p\right)} \right]
\left[f \left(E\right) - f\left(E-h\nu_p\right)\right] \Bigg]
\label{Eq:Drive_1}
\end{split}
\end{equation}
\end{widetext}
where the rate coefficient $B$ needs to be determined for low temperatures. A third term arises
in Eq.~(\ref{Eq:Drive_1})
if $h\nu_p\ge2\Delta$ ,which is not considered here since we are investigating the effect of
sub-gap photons.
Eq.~\ref{Eq:Drive_1}
 describes both absorption and emission of single photons.
 Ref.~\onlinecite{Chang_and_Scalapino_prb}
used a different rate coefficient
to describe  the interaction of microwaves at normal incidence to a superconducting film close to $T_c$
 which, in our notation, would be
$B^\prime \propto H^2 R_n/d N(0)\nu^2 $
with $H$ the magnetic field strength, $R_n$
the {\it normal-state} square sheet resistance and $d$ the film thickness .
For a SR at low temperatures  the  field does not interact with a normal-state metal in
that geometry, neither is the penetration of the field into the superconductor determined by the normal-state parameter. We take a different approach which emerges naturally by considering energy conservation.
Assuming uniform absorption, the power absorbed per unit volume of the resonator is
\begin{equation}
P_{abs}= 4N(0)  \int_\Delta^\infty dE  I_{qp}(E,\nu_p) E \rho(E,\Delta).
\label{Eq:microwave_abs}
\end{equation}
We solve Eqs.~\ref{Eq:C_and_S_1} and \ref{Eq:C_and_S_2} numerically
so that
writing $I_{qp}(E,\nu_p)=B K_{qp}(E,\nu_p)$ we can also include a power
absorption error term
\begin{equation}
\delta P =  4N(0) B  \int_\Delta^\infty dE  K_{qp}(E,\nu_p) E \rho(E,\Delta) - P_{abs},
\label{Eq:B_term}
\end{equation}
where $B$ needs to be determined to satisfy this equation, and the static non-equilibrium solution sought is $\delta P=0$.
\subsection{Quasiparticle-phonon power}
The power flow
from the quasiparticles
to the phonons  can  be found
by recognizing that the sum of the integrals on the right-hand-side of Eq.~(\ref{Eq:C_and_S_1})
 gives the
total rate of change of $f(E)$ due to  interactions with phonons,
$I_{qp-\phi}(E)$.
The energy leaving the quasiparticles per unit volume per unit time
is
\begin{equation}
P_{qp-\phi}=4 N(0) \int_\Delta^\infty dE  I_{qp-\phi}(E) E \rho(E,\Delta).
\label{Eq:qp-phonon}
\end{equation}
We
define the fractional quasiparticle-phonon power flow error term
\begin{equation}
\xi_{qp-\phi} = \frac{P_{abs}-P_{qp-\phi } }{P_{abs}}.
\label{Eq:error_qp_phonon}
\end{equation}
\subsection{Phonon cooling term}
\label{subsec:Cooling_term}
Energy is lost from a SR as non-equilibrium phonons are lost into the substrate. The
energy leaving the phonons per unit volume of the film per unit time
is given by
\begin{equation}
P_{\phi-b}= \sum_{br} N_{ion} \int_0^\infty d\Omega
D(\omega) \Omega \frac{ n(\Omega)-n(\Omega,T_b)}{\tau_{l}} .
\label{Eq:phonon_out}
\end{equation}
With a Debye model the density of states is given by
$D(\omega)=3\Omega^2/\Omega_D^3$, $\Omega_D$ is the Debye energy,
and the  sum over the phonon branches introduces an additional factor of three. We define a further error term
\begin{equation}
\xi_{\phi-b}=\frac{P_{abs}-P_{\phi-b}}{P_{abs}}.
\label{Eq:error_phonon_bath}
\end{equation}
 Eqs.~(\ref{Eq:error_qp_phonon}) and (\ref{Eq:error_phonon_bath})
 provide an important monitor of
the accuracy of the numerical solutions to the coupled equations.
\subsection{Power absorption and the effect of geometry}
Power absorption in a resonator is  dependent on its geometry. In modeling,
 we assumed that probe photons were uniformly absorbed in the SR and we ignored possible
out-diffusion of excited quasiparticles.
We assumed $P_{abs}$ was known.
In practice geometric effects  need to accounted for
 but these depend on the particular realization.
 For example  a $\lambda/4$-resonator
is a useful geometry into which to couple an external pair-breaking signal.
  The current density distribution in this device  is described by
$J(x)=J(x_0) \sin(\pi x/2 x_0)$, which is a maximum $J(x_0)$ at the shorted end $x=x_0$, and the peak power density is a factor close to  $2$-times higher than the
average.
For the same reason, as a detector the  $\lambda/4$-SR is most sensitive to changes in
the quasiparticle density  at its shorted end. Mirror currents in the ground-plane mean that the effective
volume for the power absorption  is upto twice that of the central conductor. Moreover
in a $\lambda/4$ resonator used as a detector out-diffusion of the excess quasiparticles generated by the in-coming signal
must be minimized in order to maximize the detection sensitivity, which may be achieved for example
  by using a higher energy gap
contact to the electrical ground.
In practice other resonator geometries are used and the effect of geometry and out-diffusion on power absorption can
in principle be calculated.
\subsection{Recombination times}
In what follows we calculated distribution-averaged quasiparticle recombination times $\tau_r$ for the driven $f(E)$. We used
Ref.~\onlinecite{Chang_and_Scalapino_prb} Eq.~(A9) to find the rate coefficient $R$ averaged over $f(E)$ and set
$\tau_r=(2RN_{qp})^{-1}$ with $N_{qp}=4N(0)\int_\Delta^\infty dE \rho(E) f(E)$ the non-equilibrium quasiparticle density.
If the   detected power  is small compared to the probe power this is the appropriate measure of the small-signal relaxation time.
For a thermal distribution we find    $\tau_r \equiv \langle \tau_{r}(T)\rangle_{qp}$, the  distribution-averaged recombination time described
by Kaplan {\it et al.}.
\section{Numerical parameters}
\label{sec:Numerical_parameters}
To describe  the  resonator we used material parameters appropriate for Al with $N(0)V_{BCS}=0.167$ giving
 $\Delta(0)=180\,\,{\rm \mu eV}$, $T_c=1.17\,\,{\rm K}$  and we set $T/T_b=0.1$.
We used $N(0)=1.74\times 10^{4} \,{\rm \mu eV^{-1} \mu m^{-3}}$,
$\tau_0=438\,\,{\rm ns}$ and $\tau_0^{ph}=0.26\,\,{\rm ns}$\cite{Kaplan}. The latter was calculated assuming
that the appropriate value for the phonon density of states in the calculation of $\tau_0^{ph}$ is $\alpha^2_D$
appropriate for a Debye model as given in Table~II of Ref.~\onlinecite{Kaplan}, an approach suggested in Ref.~\onlinecite{Zehnder_model}.
To be precise, in  our view the parameters needed as inputs for modeling are not collectively and with sufficient precision
known from measurement or theory.
Considering the pre-factors in Eqs.~(\ref{Eq:C_and_S_1}), (\ref{Eq:C_and_S_2}), (\ref{Eq:qp-phonon}),
and (\ref{Eq:phonon_out}),
 we found that the numerical inputs must satisfy
\begin{equation}
\frac{2\pi N(0)\tau_0^{\phi} \Delta_0\Omega_D^3}{ 9 N_{ion} \tau_0\left(k_bT_c\right)^3}=1
\label{Eq:Convergence_req}
\end{equation}
to allow a self-consistent solution
where the power errors Eqs.~(\ref{Eq:error_qp_phonon}) and (\ref{Eq:error_phonon_bath})
converged to zero.

\subsection{Parameters for a $\lambda/4$-resonator}
\label{sec:KID_model}
In  calculations  discussed later and shown in Fig.~\ref{fig:Effect_on_S21}
we investigated the effect of the driven  $f(E)$ on a representative device
modeling a $\lambda/4$ microstrip resonator as  in Ref.~\onlinecite{de_Visser}.
The complex conductivity $\sigma$ which is proportional to the normal-state conductivity $\sigma_N$
was calculated from Eqs. (3.9) and (3.10) of  Ref.~(\onlinecite{Mattis_Bardeen})
but using the non-equilibrium $f(E)$.
The surface impedance was calculated from
$\sigma$
 hence the
propagation constant and the characteristic impedance of the SR.
The modeled SR had a length of $7.6\,\,{\rm mm}$, width $3\,\,{\rm \mu m}$,  film thickness of $200\,\,{\rm nm}$,
 dielectric thickness of $200\,\,{\rm nm}$ with $\epsilon_r=3.8$ and a saturation
quality factor of $10^7$.
We set the  coupling capacitance to be $5\,\,{\rm fF}$ and
$\sigma_N=1.25\times 10^{8}\,\,{\rm (\Omega\, m)^{-1}}$, which would be typical for a clean Al film at low temperatures.
We calculated  a resonant frequency
 $\nu_0=3.934\,21\,\,{\rm GHz}$
($h\nu_0=16.2\,\,{\rm \mu eV}$)
with zero absorbed power.
\section{Numerical Method}
\label{sec:Method}
%
%
A  non-equilibrium solution for $f(E)$ and $n(\Omega)$ requires simultaneous solutions
of Eqs.~(\ref{Eq:C_and_S_1}),
(\ref{Eq:C_and_S_2}), (\ref{Eq:Non_eq_gap}),
 and
(\ref{Eq:microwave_abs}).
The task is complicated
since, for example, Eq.~(\ref{Eq:C_and_S_1}) contains terms such as
$f(E\pm\Omega)$ and likewise (\ref{Eq:B_term}) requires knowledge of
$f(E\pm h\nu_p)$,  and there are likely to be peaks in the driven distributions
arising from the high density of states of quasiparticle near $E=\Delta$.

Eqs. (\ref{Eq:C_and_S_1}), (\ref{Eq:C_and_S_2}), and (\ref{Eq:microwave_abs}) were solved using Newton's method.
We discretized the distributions $f(E)$ and $n(\Omega)$ using a $1\,\,\mu{\rm eV}$ grid with
$E_i=\Delta+i-1$, $\Omega_i=i$ and $i\in 1\dots N$ with $N=1000$
 so that that quasiparticle states  up to
 $\sim6.5\Delta(0)$ are considered.
We formed the state vector
$\bm{ \alpha } =[f_i,B, n_i]^T$
where $T$ denotes the transpose. We formed the error vector
$\bm{ \epsilon } =[{df_i}/{dt},\delta P, dn_i/dt]^T$.
The iterative procedure seeks to find $\bm{ \epsilon }^{l+1} =0 $ using
$\bm{\alpha}^{l+1}=
\bm{\alpha} ^l - \chi\left[ \bm{ J} (\bm{\alpha}^l)\right] ^{-1} \bm{\epsilon} (\bm{\alpha}^l)$
where the matrix $\bm {J} = d\epsilon_j/d\alpha_k$ is the Jacobian of the partial derivatives and $j{\rm ,\,}k\in 1\dots 2N+2$.
It is possible to derive analytical expression for the derivatives making the Jacobian efficient to evaluate.
The superscript $l$ denotes the iteration number.
 $\chi\leq 1$ is a convergence parameter  and we find
$0.8\leq\chi\leq0.95$ gives reasonably
rapid convergence typically within 10 iterations.

We assumed a starting thermal $f^0=f(E_i,T_{start})$
with an initial temperature $T_{start}\sim 2T_b$
The value chosen
for $T_{start}$ did not affect the solutions obtained merely the number
of iterations required to converge sufficiently.
 Using an earlier
estimate of the non-equilibrium  distributions reduces the number of iterations
(or increases the precision for the same computation time) and is a useful approach if
parameters such as power or phonon-trapping factors are being varied systematically.
We chose $n^0=n(\Omega_i,T_b)$
so that the phonons are initially at the bath temperature.
The aim was to find $f^l(E)$ and $n^l(\Omega)$
such that $\vert \xi_{qp-\phi}^l\vert{\rm , \,}  \vert \xi_{\phi-b}^l\vert  \le 0.1\% $ for both
power transfer error terms.
 All solutions shown below
exceed these convergence criteria in some cases by nearly an order of magnitude.

We found that in our solutions  the non-equilibrium gap calculated with Eq.~(\ref{Eq:Non_eq_gap})
changed very little from
$\Delta(T_b)$ and by a maximum of $\delta\Delta \sim 50\,\,{\rm nV}$. For this reason
we did not allow $\Delta$ to change
in the simulations. For the microwave
drive we restricted $\nu_p$ to match the discretized distributions. This means
that the onset of any photon induced peaks occurs in well-defined bins of $f(E_i)$ and $n(\Omega_i)$.

\section{Solutions for Non-Equilibrium Resonators}
\label{sec:Non-Equilbrium_solutions}
\begin{figure}
  \begin{center}
   \begin{tabular}{c}
  \psfrag{Xaxis} [] [] { ${E/\Delta}  $ }
  \psfrag{Yaxis} [] [] { ${ 10^3\,f(E)  } $  }
  \psfrag{Xaxis2} [] [] { ${E/\Delta}  $  }
    \psfrag{Yaxis2} [] [] {$  $ }
   \includegraphics[height=8.6cm, angle=-90]{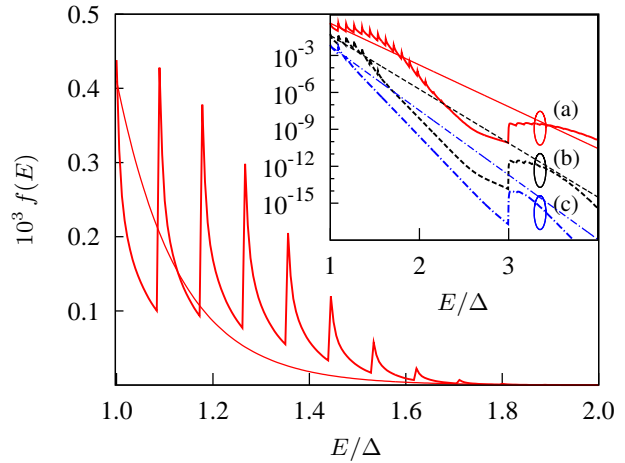}
   \end{tabular}
  \end{center}
   \caption[Fig7]
   { \label{fig:f_3_powers}
 (Color online) Non-equlibrium distribution for absorbed power $2\,\,{\rm fW/\mu m^3}$ with $T_b/T_c=0.1$  and $\tau_l/\tau_{pb}=1$.
The continuous curve is a Fermi distribution $f(E,T_N^*)$ having the same quasiparticle density.
The inset shows semi-log plots for powers of (a) $2\,\,{\rm fW/\mu m^3} (full)$, (b) $20\,\,{\rm aW/\mu m^3}$  (dashed) and (c) $0.2\,\,{\rm aW/\mu m^3}$ (dot-dash).
The associated straight lines, with the same line styles, show the
thermal distributions having the same number density of quasiparticles as the driven $f(E)$ in each case. }
   \end{figure}

In this section we show results of the modeling.
Figure~\ref{fig:f_3_powers} shows $f(E)$ with $\nu_p=3.880\,4\,\,{\rm GHz}$ ($h\nu_p=16\,\,{\rm \mu eV}$), which is
close to $\nu_0$  for the $\lambda/4$-Al SR that we  later use as an example, with
$T/T_c=0.1$, $P_{abs}=2\,\,{\rm fW/\mu m^3}$
 and
$\tau_{l}/\tau_{pb}=1$.
The multi-peaked structure is
 consistent with sequential single photon absorption;  the
drive term of Eq.~(\ref{Eq:Drive_1})  only describes single photon events.
The occurrence of this multi-peaked structure is expected if the SR is driven far from equilibrium.
Pleasingly this structure emerges  in the very first iteration of the numerical method.
In addition physically unrealistic distributions (where for example $f(E-h\nu)<f(E)$ at $E<3\Delta$) were never found.
Figure~\ref{fig:f_3_powers} includes a thermal distribution $f(E,T_{N}^*)$ where $T_N^*$ is
defined so that
$4 N(0) \int_\Delta^\infty f(E,T^*_N) \rho(E)dE = N_{qp}  $ i.e. the thermal distribution having the same number density of quasiparticles as the driven case.
The
non-equilibrium nature of $f(E)$ becomes further apparent in the
semi-log plots in the inset which are calculated for $P_{abs}$
of (a) $2\,\,{\rm fW/\mu m^3}$, (b) $20\,\,{\rm aW/\mu m^3}$ and (c) $0.2\,\,{\rm aW/\mu m^3}$ where the presence of quasiparticles with
$E\ge 3\Delta$ also
showing multiple photon induced structure can be seen.
These quasiparticles arise from absorption of $2\Delta$-phonons by quasiparticles.
A further much-reduced feature (not plotted) at $E\ge 5\Delta$ is also found.
A
recurring feature of these solutions is that the driven low-energy $f(E)$ shows
excess densities of quasiparticles at energies of order  $E < \Delta+ 10 h\nu_p$ above the equivalent
  $T^*_N$ distributions (the dashed lines).
The effect of this distortion is to increase the power carried by low energy phonons $\Omega<2\Delta$ by scattering
 compared to a thermal distribution and these phonons
are  more easily lost from the SR providing an efficient  cooling mechanism.
The distortion from the equivalent thermal distribution increases as  $P_{abs}$ increases, as does the number of photon peaks.
At energies  $E\sim 3\Delta-5h\nu_p$ the calculated $f(E)$ is increased from the by-eye straight line and at the same time
for $E\ge 3\Delta$ it is indeed the case that $f(E-h\nu)<f(E)$ for some $E$. The magnitude of both effects
are  power dependent, which
 arises from  the competing contributions of
non-equilibrium $2\Delta$-phonon re-absorption and the stimulated {\it emission} of photons inherent in Eq.~(\ref{Eq:Drive_1}).
   \begin{figure}
  \begin{center}
   \begin{tabular}{c}
  \psfrag{Xaxis} [] [] { ${\Omega/\Delta}  $ }
 \psfrag{Yaxis} [] [] { ${ P(\Omega)_{\phi-b} } (\rm{\,\, aW / \mu eV \, \mu m^3} )  $  }
  \psfrag{Xaxis2} [] [] { $ \Omega/\Delta   $  }
  \psfrag{Yaxis2} [] [] {$  $ }
   \includegraphics[height=8.6cm, angle=-90]{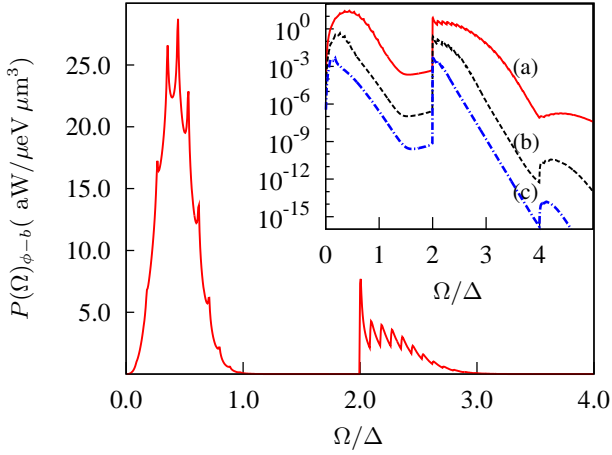}
   \end{tabular}
  \end{center}
   \caption[Fig7]
   { \label{fig:n_3_powers}
 (Color online) Contributions to the phonon power flow for absorbed power $2\,\,{\rm fW/\mu m^3}$.
The inset shows semi-log plots for absorbed powers of (a) $2\,\,{\rm fW/\mu m^3}$, (b) $20\,\,{\rm aW/\mu m^3}$
and (c) $0.2\,\,{\rm aW/\mu m^3}$. all with $\tau_l/\tau_{pb}=1$ }
   \end{figure}

Figure~\ref{fig:n_3_powers}
shows the corresponding contributions to the phonon-bath power flow integral of
 Eq.~(\ref{Eq:phonon_out}), $P(\Omega)_{\phi-b}$, for the same
drive conditions
where the presence of  non-equilibrium
$2\Delta$-phonons is seen. An additional feature at $\Omega=4\Delta$ is also found.
There are two distinct contributions to $P(\Omega)_{\phi-b}$. At low energies
we see phonons arising from the scattering of low energy quasiparticles towards the gap,
and there is structure consistent with the peaks in $f(E)$. Structure on the low energy side of the phonon
peaks is also seen which is expected as the driven $f(E)$  scatters to lower energies
and the rate of this scattering is reduced  by the occupation of the  final states by the driven
distribution itself, despite the increasing phonon density of states that would be available for the scattering to occur.
At $\Omega\ge 2\Delta$ we   see a second distinct contribution to $P(\Omega)_{\phi-b}$.
This
power  is transferred to the substrate by  pair-breaking phonons with
$\Omega\ge 2\Delta$,
which are  generated not just by recombination of the excess $f(E)$ itself but also by
 the scattering (and recombination) of those quasiparticles with $E\ge 3\Delta$.
To quantify the fraction of the power  carried by  phonons with $\Omega\ge 2\Delta$
 we  define
 $\eta_{2\Delta}=\int_{2\Delta}^\infty d\Omega (n(\Omega) -n(\Omega,T_b))/\int_0^\infty d\Omega (n(\Omega) -n(\Omega,T_b))       $.
  In the main plot of Fig.~\ref{fig:n_3_powers},
   $\eta_{2\Delta}=0.16$.
   \begin{figure}
  \begin{center}
   \begin{tabular}{c}
   \psfrag{Xaxis} [] [] { ${\Omega/\Delta}  $ }
   \psfrag{Yaxis} [] [] { ${ P(\Omega)_{\phi-b} }  \,\, (\rm{ aW / \mu eV \, \mu m^3} )  $  }
   \psfrag{Xaxis2} [] [] { $ E/\Delta   $  }
   \psfrag{Yaxis2} [] [] {$ 10^3\,f(E) $ }
   \psfrag{ABBBBB} [l] [l] {$ \tau_{l}=\tau_{pb}/2 $ }
   \psfrag{BBBBBB} [l] [l] {$ \tau_{l}=2 \tau_{pb} $ }
   \includegraphics[height=8.6cm, angle=-90]{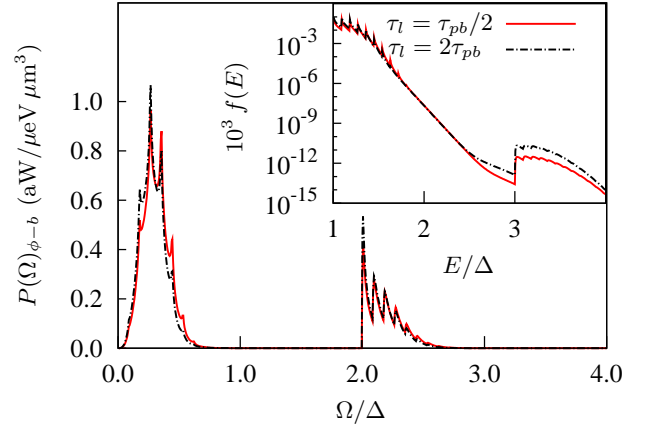}
   \end{tabular}
  \end{center}
   \caption[Fig7]
   { \label{fig:Contributions_with_trapping}
 (Color online) Contributions to the phonon power to the bath $P(\Omega)_{\phi-b} $ and inset the
associated $f(E)$ for $P_{abs}=50\,\,{\rm aW/\mu m^3}$, $T_b/T_c=0.1$, for two values of
 $\tau_{l}/\tau_{pb}=0.5$ (full lines) and $2$ (dash-dot lines)  with $\tau_l/\tau_{pb}=1$.  }
   \end{figure}
Figure~\ref{fig:Contributions_with_trapping} shows
contributions to $P(\Omega)_{\phi-b} $ for two values of $\tau_{l}/\tau_{pb}$.
 Somewhat counter-intuitively  increasing
 $\tau_{l}/\tau_{pb}$  increases the contribution
at the lowest $\Omega$  to $P(\Omega)_{\phi-b}$ whilst
simultaneously  increasing the contribution from pair-breaking phonons $\Omega\ge 2\Delta$. The
 effect on $f(E)$ is evident in the inset. As $\tau_{l}/\tau_{pb}$ is increased more
 $3\Delta$ quasiparticles are generated, these in turn generate  more pair-breaking
 phonons which can be reabsorbed before being lost from the film.
   \begin{figure}
  \begin{center}
   \begin{tabular}{c}
   \psfrag{x1} [] [] { ${\tau_{l}/\tau_{pb}}  $ }
   \psfrag{x3} [] [] { ${\tau_{l}/\tau_{pb}}  $ }
   \psfrag{y1} [t] [b] { ${ \eta_{2\Delta} }  $  }
   \psfrag{y2} [] [] { ${ T_N^*    }  (\rm{mK} ) $   }
   \psfrag{y3} [] [] { ${\tau_{r}    }  (\rm{ \mu s} ) $ }
   \psfrag{y4} [] [] { ${N_{qp}    }  (\rm{ \mu m^ {-3}} ) $ }
   \psfrag{Yaxis2} [] [] {$ f(E)\times 10^3 $ }
   \includegraphics[height=8.6cm, angle=-90]{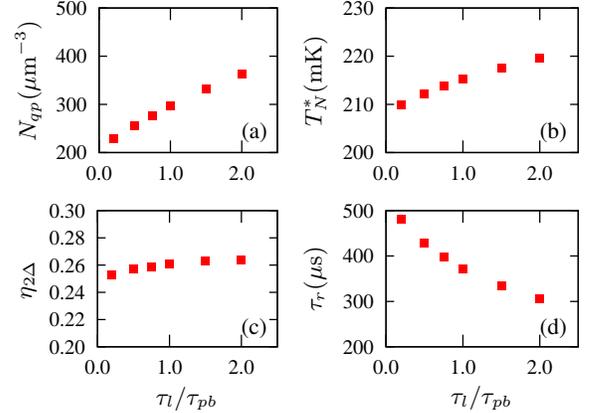}
   \end{tabular}
  \end{center}
   \caption[Fig7]
   { \label{fig:Effect_of_trapping}
 (Color online) The effect of   $P_{abs}=50\,\,{\rm aW/\mu m^3}$ as a
function of $\tau_{l}/\tau_{pb}$: (a) Quasiparticle density, (b) Effective temperature $T_N^*$,
(c) Fraction of power carried by $2\Delta$-phonons and (d) Recombination time for the
non-equilibrium $f(E)$.   }
   \end{figure}
Figure~\ref{fig:Effect_of_trapping}
shows the effect of changing $\tau_l/\tau_{pb}$ on  $N_{qp}$, $T_N^*$, $\eta_{2\Delta}$, and
$\tau_r$ for $P_{abs}=50\,\,{\rm aW/\mu m^3}$. Increasing $\tau_l/\tau_{pb}$ increases both $N_{qp}$  and  $T_N^*$
 while $\tau_r$ is reduced.
In combination this increases the relative contribution of the recombination phonons to the power flow to the bath
despite the increasing probability of re-absorption, and for this reason
$\eta_{2\Delta}$ increases somewhat. We find that for fixed power the solutions are
 related by $N_{qp}/\tau_r(1+\tau_l/\tau_{pb})=k$ where $k$ is a constant
 independent of $\eta_{2\Delta}$
 although we emphasize that  both
$N_{qp}$ and $\tau_r$ are driven non-equilibrium values.
   \begin{figure}
  \begin{center}
   \begin{tabular}{c}
   \psfrag{x1} [] [] { ${P_{abs}}(\rm{aW/\mu m^3  } )  $ }
   \psfrag{x3} [] [] { ${P_{abs}}(\rm{aW/\mu m^3  } )  $ }
   \psfrag{y1} [t] [B] { ${ \eta_{2\Delta} }  $  }
   \psfrag{y2} [] [b] { ${ T_N^*    }  (\rm{mK} ) $   }
   \psfrag{y3} [] [] { ${\tau_{r}    }  (\rm{ \mu s} ) $ }
   \psfrag{y4} [] [] { ${N_{qp}    }  (\rm{ \mu m^ {-3}} ) $ }
   \psfrag{Yaxis2} [] [] {$ f(E)\times 10^3 $ }
   \includegraphics[height=8.6cm, angle=-90]{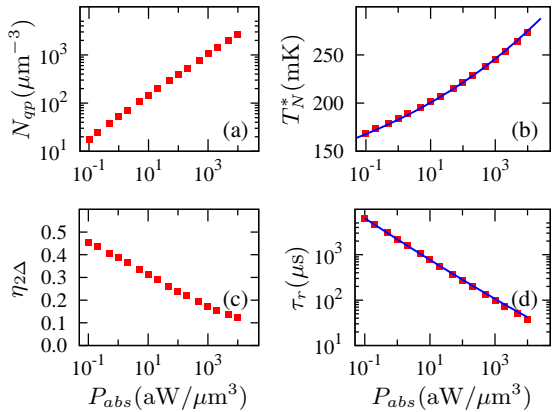}
   \end{tabular}
  \end{center}
   \caption[Fig7]
   { \label{fig:Effect_of_power}
 (Color online) The effect of  absorbed power with $\tau_l/\tau_{pb}=1$: (a) Quasiparticle density, (b) Effective temperature $T_N^*$,
(c) Fraction of power carried by $2\Delta$-phonons and (d) Recombination time for the
non-equilibrium $f(E)$. The full curve in (b) is an anayltical expression. The full curve
in (d) is the distribution-averaged thermal recombination time $\langle \tau_r(T_N^*) \rangle_{qp}$ }.
   \end{figure}

Figure~\ref{fig:Effect_of_power}
shows the effect at $T/T_c=0.1$  of varying $P_{abs}$
on  $N_{qp}$, $T_N^*$, $\eta_{2\Delta}$ and
$\tau_r$.
 Figure~\ref{fig:Effect_of_power}(a) shows that, for
all $P_{abs}$,
$N_{qp}$ exceeds the undriven thermal density calculated at $T_b$ and, to emphasize,
 for $T_b/T_c=0.1$ we calculate $N_{qp}=0.1\,\,{\rm \mu m^{-3}}$.
Readout power significantly changes the driven, static $N_{qp}$.
Figure~\ref{fig:Effect_of_power}(b) shows that $T_N^*$ is enhanced
above $T_b$ for all  read-out powers studied.
The full curve shown is an analytical expression described later in
Sec.~\ref{sec:Model}.
The distortion of $f(E)$ from even the nearest thermal distribution as $P_{abs}$ is increased
means that $\eta_{2\Delta}$ shown in Fig.~\ref{fig:Effect_of_power}(c)
is also a function of $P_{abs}$. At the lowest powers studied
$P_{abs}\sim 0.1\,\,\rm{aW/\mu m^ {-3}}$ much of the power leaving the film is carried by recombination phonons,
which is as expected given our earlier estimate showing the inefficiency of
scattering in the energy relaxation.
 As
$P_{abs}$ is increased $\eta_{2\Delta}$ is reduced and more power is carried by $\Omega<2\Delta$-phonons emitted by quasiparticle scattering. For $P_{abs}\to 0 $ we find $\eta_{2\Delta}\to 0.6 $ which interestingly is the result found for high-energy interactions
$h\nu\gg 2\Delta$.
Figure~\ref{fig:Effect_of_power}(d) shows that $\tau_r$
is reduced as  $P_{abs}$ is increased and, as expected, mirrors the increase in $N_{qp}$.
The full curve
in (d) is the distribution averaged {\it thermal} recombination time at $T_N^*$, $\langle \tau_r(T_N^*) \rangle_{qp}$,
which gives a very good description of the recombination time of the driven $f(E)$
 typically within a few $\%$. The deviation increases with
$P_{abs}$
which again is expected:  the distortion of $f(E)$ with $P_{abs}$  from the $T_N^*$ distribution increases the available final phonon densities of states
for recombination.
   \begin{figure}
  \begin{center}
   \begin{tabular}{c}
   \psfrag{x1} [] [] { ${P_{abs}}(\rm{aW/\mu m^{3}} )  $ }
   \psfrag{x3} [] [] { ${P_{abs}}(\rm{aW/\mu m^{3}} )  $ }
   \psfrag{y1} [] [] { ${ \eta_{2\Delta} }  $  }
   \psfrag{y2} [] [] { ${ T_N^*    }  (\rm{mK} ) $   }
   \psfrag{Yaxis2} [] [] { $ T_{21}^*/T_N^*  $ }
   \psfrag{Xaxis2} [] [] { ${P_{abs}}(\rm{aW/\mu m^{3}} )  $}
   \psfrag{Yaxis} [] [] {${\vert {S_{21}}\vert ^2} \,\, (\rm{dB})$ }
   \psfrag{Xaxis} [] [] {$\delta\nu\,\,(\rm{kHz}) $ }
   \includegraphics[height=8.6cm, angle=-90]{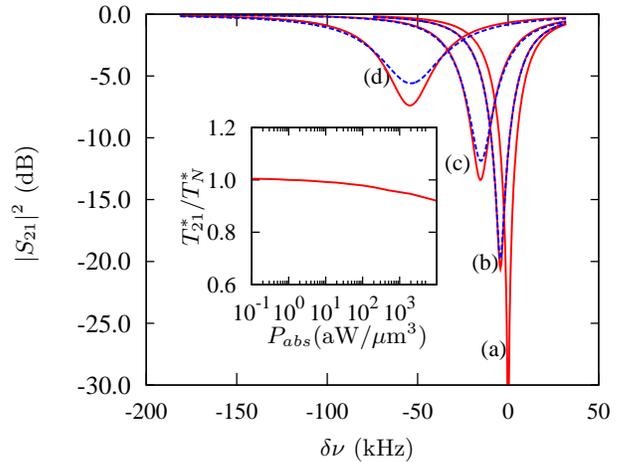}
   \end{tabular}
  \end{center}
   \caption[Fig7]
   { \label{fig:Effect_on_S21}
 (Color online) The effect of  absorbed power  on $S_{21}$ (full lines) with $\tau_l/\tau_{pb}=1$ and
 (dashed lines) the transmission calculated for the nearest thermal distribution $f(E,T_{21}^*)$
giving the same resonant frequency: (a) $P_{abs}=0$ at $T_b$,
  (b) $P_{abs}=0.1\,\,{\rm aW/\mu m^3}$,
(c) $P_{abs}=2 \,\,{\rm aW/\mu m^3}$ and (d)  $P_{abs}=50 \,\,{\rm fW/\mu m^3}$. $\delta \nu$ is referenced to
$\nu_0$ with $P_{abs}=0$.  }
   \end{figure}

Figure~\ref{fig:Effect_on_S21}
shows (full lines) calculated $\vert S_{21} \vert^2 $ for the driven $f(E)$ for  absorbed powers:
(a) $P_{abs}=0$ at $T_b$,
  (b) $P_{abs}=1 \,\,{\rm aW/\mu m^3}$,
(c) $P_{abs}=2 \,\,{\rm fW/\mu m^3}$ and (d)  $P_{abs}=50 \,\,{\rm aW/\mu m^3}$.
The dotted lines show $\vert S_{21} \vert^2 $  for the nearest
thermal distribution $f(E,T_{21}^*)$ giving the same $\nu_0$ in each case.
 A general characteristic of all
calculated transmission curves is that the driven $S_{21}$
has a higher $Q$ (it is deeper and narrower) than the
nearest $T_{21}^*$ prediction.
This arises due to occupation of final states  for absorption by  the driven distributions.
 For
increasing $P_{abs}$ the divergence increases. This is a further effect of the increasing distortion of $f(E)$
as a function of $P_{abs}$ observed in relation to
Fig.~\ref{fig:f_3_powers}. The inset shows that $T_N^*$ gives a reasonable account of $T_{21}^*$ for the range
of$P_{abs}$ considered, particularly at low powers.
   \begin{figure}
  \begin{center}
   \begin{tabular}{c}
   \psfrag{aaaa} [r] [r] { $\eta=0.59  $ }
   \psfrag{bbbb} [r] [r] { $ = 1{\,\,\,\,\,\,\,}$ }
   \psfrag{y2} [] [] { ${ T_N^*    }  (\rm{mK} ) $   }
   \psfrag{y3} [] [] { ${\tau_{r}    }  (\rm{ \mu s} ) $ }
   \psfrag{y4} [] [] { ${N_{qp}    }  (\rm{ \mu m^ {-3}} ) $ }
   \psfrag{xaxis} [] [] {${P_{abs}}(\rm{aW/\mu m^ {3}} )  $ }
   \psfrag{yaxis} [] [] {$NEP_{G-R}\,\, (10^{-19} \rm{W/\sqrt{Hz} }) $ }
	\psfrag{100} [c] [c] { $1$ }
   \includegraphics[height=8.6cm, angle=-90]{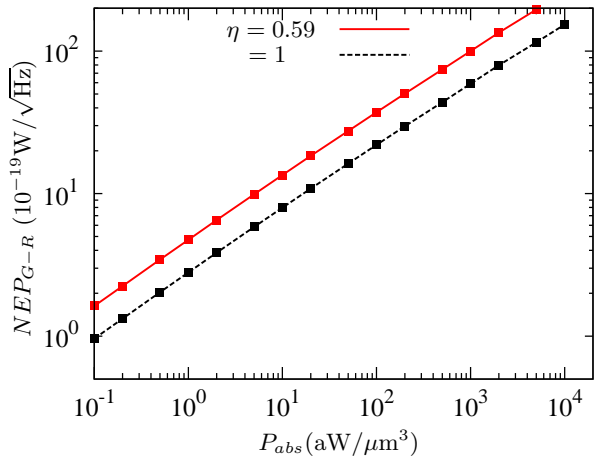}
   \end{tabular}
  \end{center}
   \caption[Fig7]
   { \label{fig:NEP}
 (Color online) The effect of  absorbed power  on generation-recombination limited NEP for 2 values of
signal detection efficiency $\eta$, resonator volume $V=1000\,\,{\rm \mu m^3}$
and $\tau_l/\tau_{pb}=1$.  }
   \end{figure}

The intrinsic limiting NEP of a superconducting detector in thermal equilibrium is determined by the random generation and recombination  of
quasiparticles.\cite{Wilson_noise} In thermal equilibrium
$NEP_{G-R}=2\Delta/\eta \sqrt{N_{qp}V/\tau_r^*}$
where $\eta$ is the fraction of {\it detected} power $P_{det}$ coupling to the
quasiparticles, $V$ is the volume of the SR,
 and the effective recombination time $\tau_r^*=\tau_r \left[ 1+ \tau_l/\tau_{pb}   \right]/2 $.
 Here the factor of two arises because, as noted by Kaplan {\it et al.} and others,
 $\tau_r$ is that of a single quasiparticle   whereas two quasiparticles are
 lost in each recombination event.\cite{Kaplan, de_Korte_SPIE, Wilson_noise_2004}
 For detection of high energy photons $h\nu_\Phi \gg 2\Delta$, $\eta\simeq0.59$.\cite{Zehnder_model, Ishibashi, Kurakado}.
 If we assume that  $P_{det}$ is small compared to
$P_{abs}$ so that $\delta N_{qp}/N_{qp}$ is small (as it must be for a linear detector) then
the relevant
 $N_{qp}$ and $\tau_r$ are as already calculated.
Figure~\ref{fig:NEP} shows $NEP_{G-R}$  as a function of  $P_{abs}$ for a SR with $V=1000\,\,{\rm \mu m^3}$
and $\tau_l/\tau_{pb}=1$ for 2 values of $\eta$, and we have assumed  that the equilibrium expression
for  $NEP_{G-R}$ applies for the driven case.
 $P_{abs}$  determines the limiting NEP and even for the
lowest $P_{abs}$ studied,  $NEP_{G-R}\sim 1-5\times10^{-19} \rm{W/\sqrt{Hz}}$.

\section{Analytical power model}
\label{sec:Model}
The preceding calculations provide insight into
 the effects of $P_{abs}$ with $\nu_p\ll 2\Delta/h$  at low reduced temperatures on low-$T_c$ SRs.
However in many situations, for example for  estimates of performance or for
 extrapolation to other materials, an
expression to approximate the key results would be extremely powerful.
Recombination determines the overall time-evolution of  the driven system,  even though we have shown that only a power-dependent fraction
$\eta_{2\Delta}$ of $P_{abs}$  is carried by the recombination phonons.
An approximate equation giving an estimate of $T_N^*$ as a function  of
$P_{abs}$
 can be derived considering energy conservation
so that
\begin{equation}
 \int_0^{P_{abs}} dP\, \eta_{2\Delta}  =   \int_{T_b}^{T^*} dT\, \frac {C_{BCS}(T)}
 { \langle \tau_r(T)^{*} \rangle _{qp} } ,
\label{Eq:P_simple}
\end{equation}
where the
denominator on the right-hand side
is the distribution-averaged  effective thermal
recombination time and
$\tau_r^*(T)=  \tau_r(T) \left[1+\tau_l/\tau_{pb} \right]/2 $.
$C_{BCS}$ is the BCS specific heat capacity which comprises two terms.\cite{Tinkham}
The first is the quasiparticle heat capacity $C_{qp}=4 N(0) d/dT\left(\int_\Delta^\infty dE E f(E) \rho(E)\right) $
 and the second arises because  the quasiparticle  energies $E$ themselves change due to their dependence
 on $\Delta$. At the (effective) temperatures considered here $d\Delta/dT\simeq0$.
The data of Fig.~\ref{fig:Effect_of_power}(c) were fitted to a  log-linear model  giving
$\eta_{2\Delta}=-0.03\ln (P_{abs}/{\rm aW\,\mu m^{-3}}) + 0.384$. At very low absorbed powers
$P\to 0$ we find $\eta_{2\Delta}\to 0.6$.
 With the same limit Eq.~(\ref{Eq:P_simple}) was solved.
 The
result is shown as the full line in Fig.~\ref{fig:Effect_of_power}~(b).
The account of  $T_N^*$ as
 a function of $P_{abs}$ is very satisfactory.

 We found that the functional form of $T_{N}^*$
  with $P_{abs}$ can be further approximated by the simpler expression
\begin{equation}\begin{split}
  P_{abs} = & \frac{1}{\eta_{2\Delta(P_{abs})}}
\Sigma_s \left( \frac {1} {1+ \tau_l / \tau_{pb}}   \right)\\
& \times \left[ T_{N}^* \exp\left( \frac {-2\Delta(T_N^*) }{k_b T_N^*}\right)    -
 T_b  \exp\left( \frac {-2\Delta(T_b) }{k_b T_b}\right)     \right] .
\label{Eq:P_very_simple}
\end{split}
\end{equation}
where $\eta_{2\Delta(P_{abs})}$ is the fraction of power carried by $2\Delta$-phonons at $P_{abs}$.
For the  Al film modeled here we found $\Sigma_s=3.4\times10^{10}\,\,{\rm W/m^3\,K} $.
This function is indistinguishable from the full  curve plotted in Fig.~\ref{fig:Effect_of_power}(b) and gives a good account of the
effect of  $P_{abs}$ on $T^*_N$ for the parameter space studied. Both
Eqs.~(\ref{Eq:P_simple}) and (\ref{Eq:P_very_simple}) provide a straightforward route to estimate $N_{qp}$ and
$\tau_r$ as a function of $P_{abs}$.

\section{Discussion and Conclusions}
\label{sec:Discussion}
We have calculated the non-equilibrium distributions of quasiparticles and phonons,
$f(E)$, $n(\Omega)$ generated by  a flux of low-energy photons $h\nu_p\ll 2\Delta$ as
a function of $P_{abs}$ for a thin-film superconducting resonator at low temperatures $T/T_c=0.1$.
Driven $f(E)$  deviate from thermal-like distributions exhibiting
structures associated with multiple probe-photon absorption and emission for all $P_{abs}$ studied.
All calculated $n(\Omega)$
 show pair-breaking phonons $\Omega\ge 2\Delta$  for all $P_{abs}$ studied.
The density of driven quasiparticles exceeds their thermal density at the bath temperature,
confirming a simple estimate based on energy conservation using thermal scattering times.
The driven $f(E)$ can be characterized in terms of an effective temperature
$T_N^*$
which also gives a good account of the distribution averaged, {\it driven} recombination time $\tau_r$
and this can be very-well approximated using simpler expressions  to calculate the {\it thermal} recombination time at $T_N^*$.
Using $N_{qp}$ and $\tau_r$ a (dark) detector Noise Equivalent Power can be  calculated.
We find that dissipation limits the achievable NEP in the range of $P_{abs}$ considered indicating
 a minimum $NEP\sim  1\times 10^{-19}\,\,{\rm W/\sqrt{Hz}}$ although we emphasize this depends on the
actual absorbed power.

 Ref.~\onlinecite{deVisser_apl_2012}
measured  a $\lambda/2$ Al resonator where we expect out-diffusion should be minimized.
That work estimated a limiting effective quasiparticle  temperature of order $160\,\,{\rm mK}$ with $T_b=100\,\,{\rm mK}$,
$N_{qp}\sim 20-70\,\,{\rm \mu m^{-3}}$,
$\tau_r\sim 3.5- 0.5\,\, {\rm \mu s} $ depending on the power, and a
 dark NEP $\sim 2\times 10^{-19}\,\,{\rm W/\sqrt{Hz}}$
at the lowest probe power. Quantifying $P_{abs}$ from the reported results is difficult without
knowing details of the embedding circuit. However our calculations shown in
Figs.~\ref{fig:Effect_of_power}(a), (b) and (d) indicate these densities, temperatures, and lifetimes would
arise for  $P_{abs}$ in the range $ 0.1-1\,\, {\rm aW/\mu m^3}$.
For the same absorbed powers, Fig.~\ref{fig:NEP} indicates a limiting dark NEP of $2-3\times 10^{-19}\,\,{\rm W/\sqrt{Hz}}$.
 The agreement with our calculations,
without free parameters, is extremely satisfactory whilst
suggesting that the  approach that we have described has merit.

In future work we will incorporate a pair-breaking source in the kinetic equations
in addition to the probe signal.  We will also
investigate the effect of the probe frequency on the driven solutions, its effect on the achievable NEP, and the scaling
of $\eta_{2\Delta}$ with material parameters.
It will also be possible to model the detection of sub-gap photons $h\nu_\Phi<2\Delta$ using a driven
resonator.
%

\end{document}